\documentstyle[multicol,aps,psfig]{revtex}
\newcommand{\bdis}{\begin{displaymath}}
\newcommand{\edis}{\end{displaymath}}
\newcommand{\bea}{\begin{eqnarray}}
\newcommand{\eea}{\end{eqnarray}}
\newcommand{\barr}{\begin{array}}
\newcommand{\earr}{\end{array}}

\begin{document}
\twocolumn[\hsize\textwidth\columnwidth\hsize\csname 
 @twocolumnfalse\endcsname

\title{ Clustering and non-gaussian behavior in granular matter }

\author{A. Puglisi$^1$,
        V. Loreto$^2$,
        U. Marini Bettolo Marconi$^{3}$,
        A. Petri$^{4}$,
        and A. Vulpiani$^{1}$}

\address{(1) Dipartimento di Fisica, Universit\`a La Sapienza,
Piazzale A. Moro 2, 00185 Roma, Italy and
 Istituto Nazionale di Fisica della Materia, Unit\`a di Roma}         

\address{(2) P.M.M.H., Ecole Sup\'erieure de Physique et Chimie
  Industrielles, 10, rue Vauquelin, 75231 Paris, France}

\address{(3) Dipartimento di Matematica e Fisica, Universit\`a di Camerino,
Via Madonna delle Carceri,I-62032 , Camerino, Italy and
Istituto Nazionale di Fisica della Materia, Unit\`a di Camerino} 

\address{(4)Istituto di Acustica O.M. Corbino,
Fossa del Cavaliere,
Consiglio Nazionale delle Ricerche, 00133 Roma}

\maketitle

\begin{abstract}

 We investigate the properties of 
a model of granular matter consisting of $N$ Brownian particles on a line
subject to inelastic mutual collisions.
This model displays a genuine thermodynamic
limit for the mean values of the
energy and the energy dissipation.
When the typical relaxation time $\tau$ associated with the 
Brownian process 
is small compared with the mean collision time $\tau_c$
the spatial density is nearly homogeneous and the velocity probability
distribution is gaussian.
In the opposite limit $\tau \gg \tau_c$ one has strong spatial
clustering, with a fractal distribution of particles,
and the  velocity probability  distribution strongly 
deviates from the gaussian one.

\end{abstract}
\vspace{0.2cm}
PACS: 81.05.Rm, 05.20.Dd, 05.40.+j
\vskip2pc]

 In the past few years granular materials have become 
an intriguing subject of research \cite{jaeger}
-\cite{mac}, since they pose
novel questions and challenges to the theorists and experimentalists.
The constituting elements of such materials are solid particles, whose
size may range from few microns to few centimeters,
and which are subject to nonconservative contact   
forces such as friction and cohesion.

 Their collective behavior is peculiar and different from other forms
of matter, such as solids, liquids or gases,  and the ordinary
statistical mechanical approach, which successfully deals with
large assemblies of microscopic particles is not adequate.

Generally speaking granular materials 
cannot be described as equilibrium systems neither from the
configurational point of view nor from the dynamical point of view.
It is known in fact that these systems remain easily trapped
in some metastable configurations which can last for long
time intervals unless they are shaken or perturbed 
\cite{nagel}.  On the other hand
while in equilibrium statistical mechanics the kinetic energy per particle
is proportional to the temperature and the velocities are gaussianly 
distributed, in the systems we consider the tails of the distribution
deviate from the Maxwell law\cite{gialli} . This phenomenon is accompanied by a
pronounced clustering of the particles
\cite{walton,goldhirsh}
or {\em inelastic collapse} \cite{mcnamara}.

Several approaches have been proposed for the study of the so-called
``granular gases''\cite{mac,kadanoff}. One crucial difference between 
ordinary gases and granular media is represented by the intrinsic 
inelasticity of the interactions among the grains, which makes 
any theory based on energy conservation , e.g. for ideal gases,
not suitable.

In the present work we study a one dimensional
mechanical model, in the spirit of the one recently introduced by Kadanoff
and coworkers \cite{kadanoff}, but containing
some important differences regarding the energy-exchange process. 
We consider $N$ identical particles 
on a circle of length $L$ \cite{diameter} obeying to the 
following equations:
\begin{equation}
\frac{d v_i}{dt}=-\frac{v_i}{\tau}+\sqrt{\frac{2 T_F}{\tau}} f_i(t)
\label{eq:eq1}
\end{equation}

\begin{equation}
\frac{dx_i}{dt}=v_i(t)
\label{eq:eq2}
\end{equation}
where, $1\leq i \leq N$, $T_F$ is the temperature of a  
microscopic medium that we discuss below, $\tau$
is the relaxation time, in absence of collisions, and $f_i(t)$ 
is a standard white noise
with
zero average and variance $<f_i(t)f_j(t')>= \delta_{ij}\delta(t-t')$. 
In addition the particles are subject to inelastic collisions
according to the rule
$v_i'-v_j'=-r(v_i-v_j)$,
where $r$ is the restitution coefficient ($r=1$ for the completely
elastic case) \cite{crucchi}.

The introduction of the viscous term takes into account
important factors, generally disregarded in simplified models, namely
the friction among particles and energy transfers among different degrees of 
freedom, which are relevants in real granular systems.
The damping and noisy terms are very natural when interactions between
particles and the environment (particle-fluid interactions) start to 
be important. Another important class of phenomena in which a viscous 
damping and a noisy term are naturally 
present is represented by the fluidized beds, where 
the vibration of the bottom of the box produces a
random force on the particles \cite{jaeger}, \cite{nagel} and \cite{olafsen}.

The model above differs from 
the one proposed by Kadanoff and coworkers because 
in the latter the particles, subject to inelastic collisions,
are confined to an interval 
of length $L$ by two asymmetric walls: the first  
reflecting them elastically and
the second supplying energy to the particles according to a gaussian
distribution at fixed temperature.
 As evident from the simulation of ref. \cite{kadanoff} 
one observes a somehow trivial
clustering of the particles next to the elastic wall. 
We found numerically a more serious shortcoming  of such model 
consisting in 
the fact that the average energy 
and the average energy 
dissipated per particle,
defined respectively as the time average of $E(t)=1/2\sum_{i=1}^N v_i(t)^2/N$
and $W(t)=(E(t_2)-E(t_1))/(t_2-t_1)$ where $t_2$ and $t_1$ represent the 
times at which two successive collisions take place,
 are not independent from the total number 
of particles, but decay exponentially as $\sim exp \,(-cN)$, showing
that the system does not possess a proper thermodynamic limit. 
The existence of thermodynamic
limit in real granular systems is not clear, but it seems 
a rather natural requirement in a statistical mechanical approach.

Due to the inelastic collisions, in order to reach a statistically 
stationary
situation some energy 
must be injected into the system. This is  achieved in our model by  
the random noise term acting on each particle.
This term mimics the action of a vibrating box. 
Notice that the boundary conditions and
the energy-pumping mechanism are different from that of 
reference \cite{kadanoff} 
and present the advantage of providing  a 'good' 
thermodynamic limits as far as the energy $E$ and energy dissipation
$W$ are concerned,  i.e. $<E>$ and $<W>$ become independent of $N$ for 
large values of $N$. 

Moreover our system does not have walls and the clustering is non-trivial. 
On the other hand, the energy feeding mechanism adopted in \cite{mac}
forces to introduce a somehow artificial cooling of the particles,
by renormalizing the velocity of the center of mass at each time step.
In our formulation this procedure is overcome by the presence of 
the thermal bath.

For each given choice of $r$ and $\tau$ the system, after a long
transient, reaches a stationary state with certain properties.
The presence of two time scales, namely $\tau$ and the mean collision
time $\tau_c$, leads to different dynamical regimes. 

{\bf a}) When $\tau \ll \tau_c$ it is easy to argue that the grains reach 
a rather simple statistical equilibrium
and that their velocity distribution is that of an ideal gas 
with an effective temperature $T_F^*$, slightly lower than the 
temperature of the 'heat bath', $T_F$.  

{\bf b}) In the opposite limit $\tau \gg \tau_c$ the driving mechanism towards
the macroscopic stationary state is 
dominated by the collision process itself.

Two phenomena are observed:

1) the velocity distribution ceases to be gaussian and the deviation
becomes more and more pronounced with decreasing values of the restitution
coefficient $r$.

2) The spatial distribution becomes strongly inhomogeneous. 

\begin{figure}
\centerline{
\psfig{figure=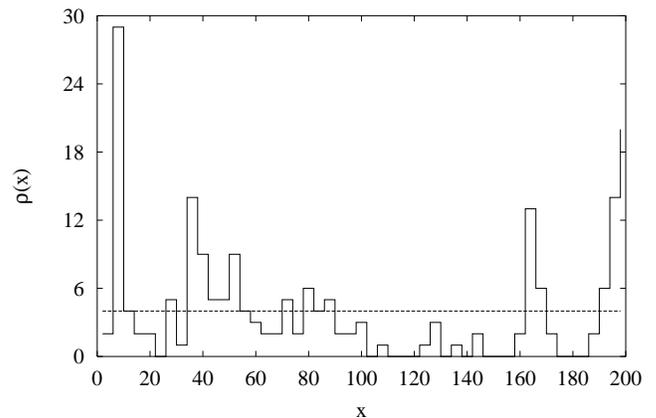,width=8.5cm,angle=-90}}
\caption{Snapshot of the particle density 
at a given time for $T_F=1$, $r=0.6$, 
$\tau=100$, $N=200$, $L=200$. All quantities are in arbitrary units. 
The dashed line represents the homogeneous density.}
\label{snapshot}
\end{figure}

\begin{figure}
\centerline{
\psfig{figure=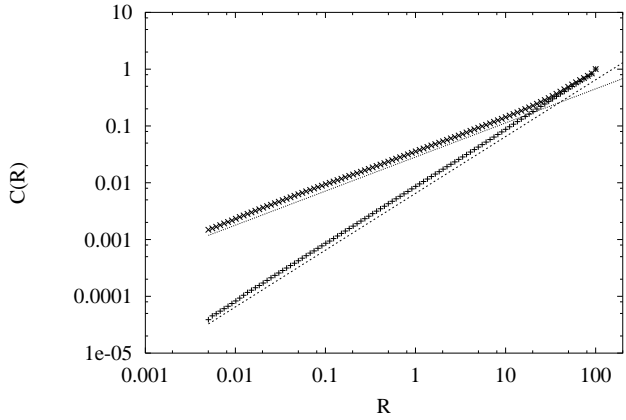,width=8.5cm,angle=-90}}
\caption{$C(R)$ against $R$ 
for two different choices of parameters: $\tau=100$, $r=0.6$ (top) and
 $\tau=100$, $r=0.99$ (bottom). In both cases $T_F=1$, $N=200$ and $L=200$.
The dimension takes respectively
the values  $d_2=0.59$  and $d_2=1$.}
\label{gr-dim}
\end{figure}

It is worth to stress that, at variance with the clusterization 
in Ref.\cite{kadanoff}, in our case the clusters are created and 
destroyed continuously in the system as long as the system evolves. 
The inhomogeneity in the spatial distribution
of the grains, see Fig.(\ref{snapshot}), can be quantitatively characterized 
by the so called Grassberger-Procaccia dimension
$d_2$ that we compute from the correlation function $C(R)$ defined as:
\begin{equation}
C(R)=\frac{1}{N^2}
\frac{1}{T}\int_0^{T} dt \sum_{i<j}\theta(R-|x(t)_i-x_j(t)|)\sim R^{d_2}
\label{eq:eq4}
\end{equation}
where $T$ represents the duration of the simulation.
In  Fig.(\ref{gr-dim}) we show $C(R)$ {\it vs} $R$ for a clustering
situation.
It turns out that $d_2$ is lower than
$1$ when $r<1$, e.g. for $\tau=100$ and $r=0.6$, $d_2=0.59$,
while for $\tau=100$ and $r=0.9$ $d_2\simeq 1$.

Notice that in this case  the stationary regime 
is brought about by the collisions and that these occur more 
frequently in the regions of higher density.
For instance for $\tau=100$ and $r=0.7$ one gets for the number of
collisions as a function of the spatial density $\rho$, 
$N_{coll} (\rho) \sim \rho^2$.

Assuming that in the stationary state
the power dissipated through the collisions must balance the power adsorbed 
from the heat bath, one derives the following relation between the temperature
$T_F$, the average energy $E$, the power dissipated by the collisions $W$ and $\tau$ 
\cite{puglisi}:

\begin{equation}
W=\frac{1}{\tau}(T_F- 2 E)
\label{eq:eq3}
\end{equation}
Notice that in the absence of collisions (or $\tau \ll \tau_c$)
$T_F= 2 E$ and $W=0$ as in the ideal gas. 
Eq. (\ref{eq:eq3}) is well satisfied
for different values of $\tau$ and $r$.

\begin{figure}
\centerline{
\psfig{figure=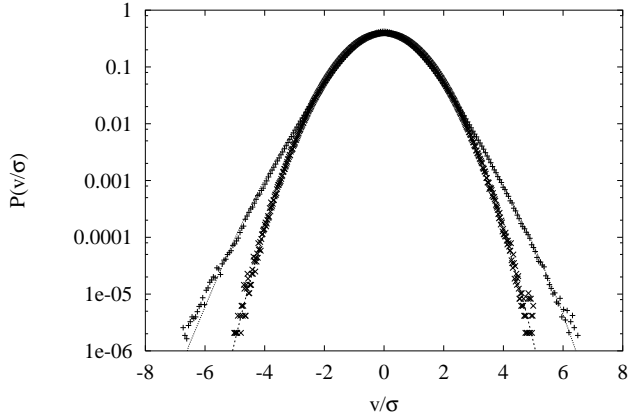,width=8.5cm,angle=-90}}
\caption{Rescaled velocity distribution
$P(v/\sigma)$ against $v/\sigma$: crosses
  are simulation data with $\tau=100$, $r=0.7$, the dashed line represents
  toy-model data, X's are simulation data with $\tau=0.01$
  and $r=0.99$ and the dot-dashed line represents the gaussian distribution.
In both cases $T_F=1$, $N=200$ and $L=200$.}
\label{pv-modello}
\end{figure}

As we quoted already in the inelastic regime one observes a strong
deviation from the gaussian behavior for the velocity distribution. 
In Fig.(\ref{pv-modello}) we display the velocity distribution in a
nearly elastic case ($\tau=0.01$ and $r=.99$) and in a strong 
inelastic regime ($\tau=100$ and $r=0.7$). As it is possible to see 
it exists an evident departure from the gaussian behavior
in the inelastic case, where the velocity distribution shows
almost exponential tails. The above result is close to that
observed in refs. \cite{gialli}, \cite{olafsen}. 

Let us now try to relate the clustering properties of the system to
the velocity distribution. In order to do that we consider the
following quantities: the distribution of boxes, $N_{box} (m)$,
containing a given
number, $m$, of particles  and the average kinetic energy, $E_{kin} (m)$, 
in a box occupied by $m$ particles \cite{notascatole}. Making the hypothesis
that in each box the average velocity of the particles is zero, 
i.e. $<v(m)>=0$ (very well confirmed by the numerical data), 
one finds that $E_{kin}(m)$ provides a measure of the

 variance of the velocity distribution in each box: 
$E_{kin}(m) \simeq \frac{1}{2} <v^2(m)> = \frac{1}{2} \sigma^2(m)$.
We consider first the non-clusterized case 
($\tau << 1$ and $r \simeq 1$). Within 
 this regime we find from the simulations that:
\begin{eqnarray}
\sigma_{elas}^{2}(m) \simeq  const. \,\,\,\,\,\,
N_{box}^{elas}(m) =  \frac{\lambda^{m} e^{- m} }{ m! },  
\label{gauss}  
\end{eqnarray}
where $\lambda= N/N_{discr}$ is the average number of 
particles in each box and $N_{box}^{elas}(m)$ is a Poisson distribution.
 By assuming in each box a gaussian
velocity distribution with a constant variance $\sigma_{elas}^{2}(m)$
it turns out that the global velocity distribution $P_{elas} (v)$
is gaussian.  Let us recall that the Poisson
distribution is the one associated with a process  
of putting independently $\lambda N$ particles into $N$ boxes.

Let us turn to the non-elastic case. If 
$\tau=100$ and $r=0.7$, considering the occupied 
boxes ($m > 0$),
we obtain from the simulations: 
\begin{eqnarray}
\sigma_{inel}^{2}(m) \sim  m^{-\beta} \,\,\,\,\,\,
N_{box}^{inel}(m) =  \frac{e^{- \alpha m} }{ m },  
\label{inel}  
\end{eqnarray}
with $\beta\simeq 0.5$ and $\alpha\simeq 0.14$.
Let us compute from these scalings the global velocity distribution. 
Taking into account that the spatial probability distribution of the particles 
is $N_{box}^{inel}(m)$ and assuming that their local
velocity distribution is gaussian, but
with a variance $\sigma_{inel}^{2}(m) \simeq  
m^{-\beta}$ which depends on the occupancy,
we obtain, for the global velocity distribution $P_{inel}(v)$,
which in the continuum limit should correspond to
 \begin{equation}
P_{inel}(v) \simeq \sum_{m=1}^{\infty} e^{(- \frac{v^2 m^{\beta}}{2})} 
e^{-\alpha m}.
\end{equation}
 We stress how it exists an
astonishing agreement between the numerical results and the ones
obtained with a toy model which just makes the following hypothesis:

{\bf i)} non-Poissonian distribution for the box occupancy; 

{\bf ii)} gaussian distribution of velocities in each box with
a density-dependent variance.

The hypothesis about the scaling relation between the velocity
variance and the local density, apart from being justified numerically,
can be understood in the following way.
The stationarity and the scale-invariance of the cluster distribution,
implies a certain distribution of lifetimes for the clusters. In particular 
each cluster has a lifetime which is inversely proportional
to its size. The scale-invariant cluster-size distribution
thus implies a scale-invariant distribution for the lifetimes.
The cluster lifetime (its stability) is strictly related to the variance
of the velocity distribution inside the cluster itself. In order to ensure the stability of a
cluster in a stationary state we have to require that the velocities 
of the particles belonging to it are not too different, or equivalently
that the variance of the distribution is smaller the higher 
the density. So, given a 
scale-invariant distribution of clusters one would expect a scale-invariant distribution 
of variances (\ref{inel}).
An independent check is provided by the behavior of the average relative
velocities in boxes with different numbers of particles. 
For $\tau=100$ and $r=0.7$ one obtains:
\begin{equation}
v_{rel}^{inel} (m) \sim m^{-\gamma}
\end{equation}
with $\gamma \simeq 0.3$, which indicates that the stability of a 
cluster is connected with the smallness of the velocity 
fluctuations inside it.
In other words, the spatial clusterization corresponds 
to a clusterization in velocity space.

This analysis shows that in the
present model it does exist 
a relation between the clustering phenomenon 
(in physical space or in velocity space) 
and the non-gaussian distribution of the velocities. 

A natural question to ask is whether our findings can be an
artifact of the one dimensional dynamics. Can one expect to
observe the non-Maxwell distribution and the clusterization even in 
higher dimensions? 
In order to clarify such issue, we consider a different 
stochastic process.
All the particles perform the  Brownian motion 
described by eq.(\ref{eq:eq1}) 
if $t \epsilon [K \Delta t, (K+1) \Delta t]$,
(where $K$ is an integer number) while at the instants
$t_K=K \Delta t$ each particle may collide with probability $p$
with one of those particles which are spatially close to it
according to the
Boltzmann {\it Stosszahlansatz} (BS).

In practice we perform the following algorithm {\em \`a la} Bird\cite{Bird}:
at each discrete time $t_K$ for each particle, $i$, we extract 
 out of a uniform distribution
in the interval $[0,1]$ a random number $y$. If $y>p$ there is
no collision, otherwise the particle $i$ scatters with another particle $j$
if $|x_i(t_K)-x_j(t_K)|<l$, and their collision probability 
is proportional to  $|v_i(t_K)-v_j(t_K)|$. 
This process renders the BS approximation exact; in fact in the limit 
$N\to \infty$, $p \to 0$, $l \to 0$,
$\Delta t \to 0$ one can write for the stochastic process described
above the following Boltzmann equation for the one particle 
distribution $P(x,v,t)$ \cite{wag}:
\begin{equation}
\frac{\partial}{\partial t} P
+\frac{\partial}{\partial x} (vP)-\frac{1}{\tau}
\frac{\partial}{\partial v} (vP)-\frac{T_F}{\tau}
\frac{\partial^2}{\partial v^2}P= 
\frac{\partial}{\partial t}P|_{coll}
\label{eq:eq5}
\end{equation}

\begin{eqnarray}
\frac{\partial}{\partial t}P|_{coll}=& \frac{ 4 \Lambda} {(1+r)^2}  
 \int dv' P(x,v',t) 
P(x,\frac{(2v-(1-r)v')}{(1+r)},t)|v'-v|  \nonumber \\ 
& - \Lambda \int dv' P(x,v',t) P(x,v,t)|v'-v|
\end{eqnarray}

where $\Lambda \sim p/\Delta t \sim 1/\tau_c$ and depends on the
mean particle density.

It is easy to show that in the limit of elastic collisions $(r=1)$
the velocity distribution becomes gaussian for any value of $\tau$,
$\Delta t$ and $p$. On the contrary for $r \neq 1$ an analytical
solution is not known.

In order to further clarify the relevance of the dimensionality, we
have performed some simulations in $2d$. In this case we find 
non gaussian tails for the velocity distribution, together with clusterization,
at large values of $\tau$ and strong inelasticity. 
In $2d$ the velocities in the collision change according to the following rule:
\begin{equation}
{\bbox v_i}(t_K+0^+)-{\bbox v_j}(t_K+0^+)
=r \hat e({\bbox v_i}(t_K)-{\bbox v_j}(t_K))
\end{equation}
where $\hat e$ is a two-dimensional unit vector of random orientation. 
In this case the results are qualitatively similar to the ones 
obtained in one dimension, i.e. at large values of $\tau$ and strong 
inelasticity the Grassberger-Procaccia dimension is smaller than $2$
and the velocity distribution is not gaussian.


In summary we introduced a model of granular gas with inelastic
collisions between the particles. The system exhibits a variety of
regimes ranging from a completely elastic case without clusterization 
and gaussian distribution for the velocities to an inelastic regime
with strong clusterization and non-gaussian velocity 
distribution. In this framework we have shown a possible scenario
to relate the clustering properties of the system to
the velocity distributions.
These results seem promising and give us the hope to be able to
perform analytic work based on the Boltzmann approach in order to
clarify further the model.



\begin{references}


\bibitem{jaeger} H.M. Jaeger and S.R. Nagel, {\em Science} {\bf 255},
  1523 (1992);  H.M. Jaeger, S.R. Nagel and R.P. Behringer, {\em Phys. Today}
  April 1996.

\bibitem{nagel} for a general overview see: 
H.M. Jaeger, S.R. Nagel and R.P. Behringer, {\em Rev. Mod. Phys.} 
{\bf 68}, 1259 (1996) and references therein.

\bibitem{walton} O.R. Walton, in {\em Particulate Two-Phase Flow} 
Part I, edited by M.C. Roco (Butterworth-Heinemann, Boston), p.884 
(1992).

\bibitem{goldhirsh} I. Goldhirsch and G. Zanetti, {\em
    Phys. Rev. Lett.} {\bf 70}, 1619 (1993).

\bibitem{hopkins} M.A. Hopkins and M.Y. Louge, {\em Phys. Fluids A}
 {\bf 3}, 4 (1990).

\bibitem{mcnamara} S. McNamara and W.R. Young, {\em Phys. Fluids A}
 {\bf 4}, 496 (1992).

\bibitem{mac}
D.R.M. Williams, F.C. MacKintosh, {\em Phys. Rev. E.} {\bf 54}, R9
(1996).

\bibitem{gialli}
Y. H. Taguchi and H. Takayasu, Europhys. Lett. {\bf 30}, 499 (1995)


\bibitem{kadanoff}
Y.Du, H.Li, L.P. Kadanoff, {\em Phys. Rev. Lett.} {\bf 74}, 1268 (1995);
K. Geisshirt, P. Padilla, E. Pr{\ae}stgaard and S. Toxvaerd, {\em Phys. Rev. E}
{\bf 57}, 1929 (1998). 


\bibitem{diameter}
In one dimension we do not need to
consider explicitly the diameter of the particles, which can be accounted for
by a rescaling of the system size $L$. 

\bibitem{crucchi}
A similar model has been discussed by T. Reichmuller, L. Schimansky-Geier,
D. Rosenkrantz and T. Poschel 
J. Stat. Phys. {\bf 86}, 421 (1997).
However, in this work the collision rules are chosen according to the
"follow the leader" prescription, suggested by the traffic problem.

\bibitem{olafsen} J.S.Olafsen and J.S.Urbach, cond-mat/9807148

\bibitem{puglisi} A. Puglisi, Tesi di Laurea at the University of Rome
  {\em La Sapienza} (1997).

\bibitem{notascatole} In order to measure these quantities we
considered the system as divided in $N_{discr}=100$ boxes.
We have checked that the results are not dependent, except for
finite-size effects, on the discretization used.

\bibitem{Bird} 
G.A. Bird, Phys. Fluids {\bf 13}, 2676 (1970).

\bibitem{wag} W. Wagner, {\em J. Stat. Phys.} {\bf 66}, 1011 (1992) and
C. Cercignani, R. Illner, M. Pulvirenti, 
"The Mathematical Theory of Dilute Gases" Springer Verlag (1994).

\end{references}
\end{document}